\pdfoutput=1

\documentclass[twocolumn,aps,pra,superscriptaddress]{revtex4}

\usepackage{times}
\usepackage{graphicx}
\usepackage{epsfig}
\usepackage{amsfonts}
\usepackage{amsmath}
\usepackage{amssymb}
\usepackage{color}
\usepackage{multirow}
\usepackage[colorlinks=true,citecolor=blue,urlcolor=blue]{hyperref}
\usepackage{float}

\def\endproof{\vrule height6pt width6pt depth0pt}

\begin{document}


\title{Simple method for experimentally testing any form of quantum contextuality}


\date{\today}


\author{Ad\'{a}n~Cabello}
\affiliation{Departamento de F\'{\i}sica Aplicada II, Universidad de Sevilla, E-41012 Sevilla, Spain}


\begin{abstract}
 Contextuality provides a unifying paradigm for nonclassical aspects of quantum probabilities and resources of quantum information. Unfortunately, most forms of quantum contextuality remain experimentally unexplored due to the difficulty of performing sequences of projective measurements on individual quantum systems. Here we show that two-point correlations between binary compatible observables are sufficient to reveal any form of contextuality. This allows us to design simple experiments that are more robust against imperfections and easier to analyze, thus opening the door for observing interesting forms of contextuality, including those requiring quantum systems of high dimensions. In addition, it allows us to connect contextuality to communication complexity scenarios and reformulate a recent result relating contextuality and quantum computation.
\end{abstract}




\maketitle


\section{Introduction}


Quantum theory (QT) is in conflict with noncontextual realism \cite{Specker60,Bell66,KS67}, that is, with the assumption that the results of measurements reveal pre-existing properties that are not affected by compatible measurements. Noncontextual realism is a legitimate assumption in so far as the statistics of the measurement outcomes are not perturbed by other measurements, or, in other words, whenever measurements cannot be used to communicate information. No signaling holds when the measurements are spacelike separated. If this is the case, noncontextual realism is called local realism. It is well known that QT is also in conflict with local realism \cite{Bell64}.

However, most quantum violations of noncontextual realism do not occur in scenarios in which parties perform spacelike separated measurements. Still, they can be reveled in experiments \cite{Cabello08}. This shows that contextuality, besides being ``a class of `paradoxes' which result from counterfactual logic'' \cite{Peres02}, is a phenomenon that can be experimentally tested \cite{Szangolies15}.

Among the violations of noncontextual realism that cannot be observed in Bell-like tests there are many cases of interest: Qutrit contextuality \cite{KCBS08}, quantum-state-independent contextuality \cite{Cabello08,BBCP09,YO12}, simple fully contextual correlations \cite{Cabello13b}, contextuality associated to universal quantum computation via magic states \cite{HWVE14,DGBR15,RBDOB15}, and absolute maximal contextuality \cite{ATC15}. This variety of examples motivates the challenge of whether there is a unified universal way to test any form of contextuality.

Certifying contextuality requires observing the violation of a noncontextuality (NC) inequality \cite{KCBS08,Cabello08}, which is an inequality for correlations satisfied by any noncontextual hidden-variable theory (NCHVT), i.e., for any theory assuming noncontextual realism. Testing a NC inequality entails considerable difficulty and requires solving some problems.


\begin{figure}[tb]
\vspace{-2.4cm}
 \includegraphics[width=0.48 \textwidth]{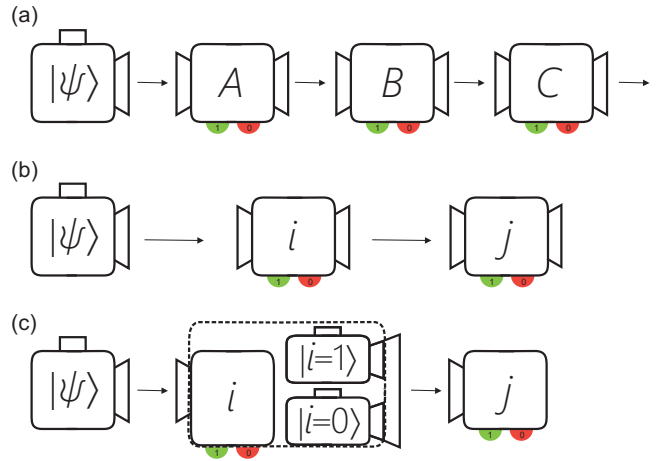}
 \vspace{-2.6cm}
 \caption{(color online). (a) A typical contextuality experiment requires preparing individual quantum systems in a quantum state $|\psi\rangle$ and performing a sequence of projective measurements on each of them. (b) Contextuality experiments can be substantially simplified by exploiting that any form of contextuality can be tested by measuring sequences of only two compatible binary observables. The problem is that in most quantum systems sequential projective measurements are unfeasible. In these cases, a possible test of contextuality (not without drawbacks) is one based on the scheme in panel (c). It requires replacing the projective measurement of $i$ in panel (b) by a demolition measurement of $i$ followed, in those trials with outcome $i=1$, by a preparation of the eigenstate of $i$ with eigenvalue $1$ or, in those trials with outcome $i=0$, by a preparation of the quantum state predicted by L\"uders's rule. \label{Fig1}}
\end{figure}


The first difficulty is related to the fact that in a test of a NC inequality every observable must be measured using the same device in any context. An alternative would be, before conducting the experiment, certifying that different devices measure the same observable since they produce the same distributions of probability for any possible state. However, this is unfeasible using a finite sample of states. Without such certification it seems unreasonable to assume that measurement outcomes do not depend on the context when measurements themselves do \cite{ABBCGKLW13}. This problem can be avoided by designing devices that exactly measure each of the required observables in any possible context.

The second difficulty comes from the fact that, except for the case of Bell inequalities (which are NC inequalities), testing other NC inequalities requires performing sequential measurements of quantum observables \cite{LKGC11}, as illustrated in Fig.\ \ref{Fig1} (a). This is challenging. Recall that the measurement of a quantum observable, as defined by von Neumann \cite{vonNeumann32} (i.e., a projective measurement), should not only produce an outcome (that should be ``indelibly recorded'' \cite{MW84} before any further measurement) but also transform the quantum state according to L\"uders's rule \cite{Luders51}. The problem is that only a few experiments have successfully implemented sequential projective measurements (e.g., Ref.\ \cite{KZGKGCBR09}). The typical measurements on quantum systems are actually demolition measurements which absorb the system, preventing subsequent measurements, or fail to transform the state of the system according to L\"uders's rule.

Nevertheless, the main difficulty is achieving the no-signaling condition with sequential measurements. Recall that in a contextuality test each measurement should not affect the probabilities of the outcomes of any subsequent measurement. That is, if the correlations between $A$ and $B$ and between $A'$ and $B$ both appear in the NC inequality and $B$ is measured in the second place, then $\varepsilon \equiv |\sum_a P(a,b|A,B)-\sum_{a'} P(a',b|A',B)|$ must be zero for any $b$. The problem is that it is impossible to have $\varepsilon=0$ in actual experiments since the number of trials is finite, long-lasting experiments suffer drifts, and there are inevitable imperfections \cite{Szangolies15,GKCLKZGR10}. One could argue that the condition of perfect no signaling is not even achieved in experiments with spacelike separated measurements (e.g., $\varepsilon = 0.07 \pm 0.10$ in Ref.\ \cite{Hensen15}). However, in these experiments no signaling can be assumed on physical grounds and a nonzero $\varepsilon$ can be attributed to statistical fluctuations. In contrast, in experiments with sequential measurements, a nonzero $\varepsilon$ should be taken into account in the analysis \cite{Szangolies15,Winter14,DKL15}. A practical issue, however, is that these analysis are in general difficult or are not well defined beyond the case of two-point correlations.

Despite these problems, little effort has been made for designing contextuality tests with small $\varepsilon$ that are easy to analyze. From this perspective, a main challenge is to have tests in which the influence of past measurements on future measurements is experimentally indistinguishable from the influence of future measurements on past measurements. More precisely, experiments in which $\varepsilon$ is zero within the same experimental error that $\varepsilon' \equiv |\sum_b P(a,b|A,B)-\sum_{b'} P(a,b'|A,B')|$. Assuming causality, $\varepsilon'$ is, by definition, zero. However, when calculated from experimental data, this zero will be inevitably affected by an error. Only if the experimental value of $\varepsilon$ is compatible with zero and is affected by a similar error can one reasonably assume that no signaling is satisfied within the experimental precision. In this case, the violation of the NC inequality can be taken as a compelling evidence of contextuality.

Another difficulty comes from the infeasibility of implementing projective sequential measurements on most quantum systems, particularly on high-dimensional quantum systems \cite{MTT07,PB11,DBSBLC14,CEGSXLC14,CAEGCXL14} which are necessary for some forms of quantum contextuality \cite{Cabello08,Cabello13b,ATC15,CDLP13}. One solution is implementing sequentially the unitary transformations corresponding to the desired measurements and detect the systems afterwards. This is the approach adopted in some experiments with photons \cite{MWZ00,LLSNRWZ11}, neutrons \cite{HLBBR03,BKSSCRH09}, and molecular nuclear spins \cite{MRCL09}. Other solution, closer to the ideal contextuality experiment in Fig.\ \ref{Fig1} (a), consists of encoding the outcome of each measurement in an extra degree of freedom (e.g., extra spatial modes \cite{ARBC09}, time bin \cite{AACB13}, or polarization \cite{MANCB14}) before performing the next one. However, this solution has one disadvantage: (I) The complexity of the setup grows exponentially with the number of sequential measurements. This does not only entail practical problems but also introduces a conceptual issue since these extra degrees of freedom could provide the memory classical systems need to simulate quantum contextuality \cite{KGPLC11,FBVFLLFC15}. Moreover, both solutions have other disadvantage: (II) The observers cannot delay the choice of the next measurement to the moment in which the result of the previous measurement is indelibly recorded. As a result, the system is not really forced to answer before the context is fixed \cite{PR88}. This indicates that other interesting challenge is designing contextuality tests for high dimensional quantum systems and free of disadvantages (I) and (II).

In this paper we show how to reveal any form of quantum contextuality in an experiment requiring sequences of only two measurements and allowing a better control on the experimental imperfections and a simple analysis. In addition, we show how to design contextuality tests for quantum systems in which sequential projective measurements are unfeasible and free of disadvantages (I) and (II).


\section{Main result}


In this section we show that two-point correlation experiments between binary compatible observables are sufficient to test any form of quantum contextuality. In the next section we will explain why this is important.

To understand what we mean by ``any form of quantum contextuality,'' we start by recalling that any given NC inequality can be associated to graph \cite{CSW14}. This association requires converting the linear combination of probabilities in the initial NC inequality into a positive linear combination $S$ of probabilities of yes-no tests (represented in QT by rank 1 projectors). $S$ can then be associated to a graph $G$ in which each probability is represented by a vertex while edges connect vertices corresponding to probabilities of exclusive events (i.e., corresponding to alternative outcomes of a sharp measurement \cite{CY14}). The maximum of $S$ for NCHVTs and QT are two characteristic numbers of $G$: The independence number, $\alpha(G)$, and the Lov\'asz number, $\vartheta(G)$, respectively \cite{CSW14}. Reciprocally, any graph $G$ can be converted into a NC inequality such that the bound for NCHVTs is $\alpha(G)$ and the maximum in QT is $\vartheta(G)$ \cite{CSW14}. This establishes a one-to-one connection between any possible quantum violation of any possible NC inequality and a graph. It is in this sense that any form of quantum contextuality can be represented by a graph.

Here we show that any NC inequality represented by a graph $G$ can be further transformed into a NC inequality involving only two-point correlations, while still having $\alpha(G)$ as upper bound for NCHVTs and $\vartheta(G)$ as maximum in QT.

{\em Theorem.} For any graph $G$ with vertex set $V(G)$ and edge set $E(G)$ the following inequalities are tight:
\begin{equation}
\begin{split}
\label{main}
{\cal S} \equiv
\sum_{i \in V(G)} P(1|i) - \sum_{(i,j) \in E(G)} P(1,1|i,j)
& \stackrel{\mbox{\tiny{NCHVTs}}}{\leq} \alpha(G) \\
& \stackrel{\mbox{\tiny{QT}}}{\leq} \vartheta(G),
\end{split}
\end{equation}
where $P(1|i)$ is the probability of obtaining result $1$ when observable $i$ is measured and $P(1,1|i,j)$ is the joint probability of obtaining result $1$ for $i$ and result $1$ for $j$.

The quantum maximum $\vartheta(G)$ can be attained by preparing a quantum state $|\psi\rangle$ and measuring a set of rank 1 projectors $\{i=|i\rangle \langle i|\}$ such that each projector is associated to a vertex of $G$, adjacent vertices are associated orthogonal projectors, and $\sum_i |\langle i | \psi \rangle |^2 = \vartheta(G)$. The set $|\psi \rangle \cup \{|i\rangle\}$ is called a Lov\'asz-optimum-orthogonal representation of $\overline{G}$ \cite{CDLP13}. Its existence is guaranteed by the definition of $\vartheta(G)$.


\begin{figure}[tb!]
\includegraphics[trim = 3.6cm 5.2cm 3.6cm 5.4cm,clip,width=8cm]{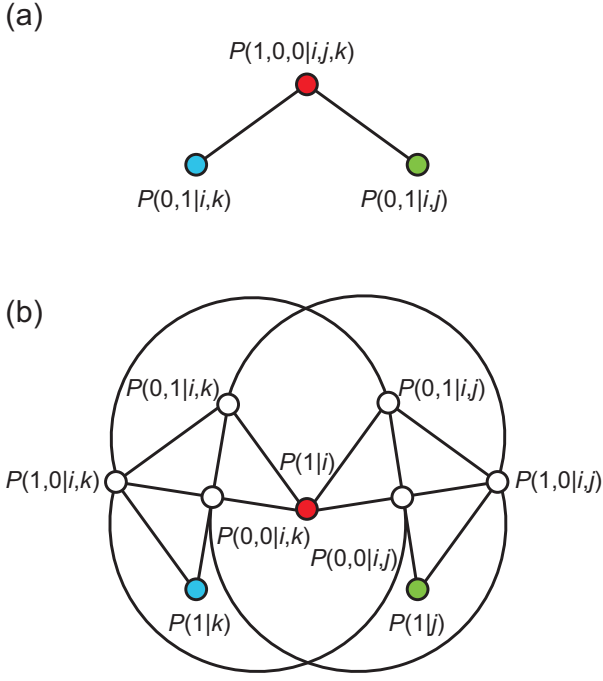}
 \caption{(color online). Simplest nontrivial example of how the graph $G$ is related to the graph $G'$. (a) For any graph $G$ with vertex set $V(G)$, there is a NC inequality of the type $S = \sum_{i \in V} P(1,0,\ldots,0|i,i_1,\ldots,i_{n(i)}) \stackrel{\mbox{\tiny{NCHVTs}}}{\leq} \alpha(G)$, which is violated by QT up to $\vartheta(G)$; the set $\{i_j\}_{j=1}^{n(i)}$ contains the observables corresponding to vertices adjacent to vertex $i$. The probabilities in this inequality are well defined both in NCHVTs and in QT but, in general, involve incompatible tests. This is a problem for designing experiments. (b) By replacing the probabilities $P(1,0,\ldots,0|i,i_1,\ldots,i_{n(i)})$ in $G$ by $P(1|i)$ and every edge $(i,j) \in G$ by the vertices corresponding to $P(0,0|i,j)$, $P(0,1|i,j)$, and $P(1,0|i,j)$ and adding edges between vertices corresponding to exclusive events, we construct a new graph $G'$ which is associated to a new NC inequality, given in Eq.\ (\ref{main}), which has the same bounds for NCHVTs and QT that the NC inequality associated to $G$, but involves only compatible observables and can be tested by measuring only two-point correlations. \label{Fig2}}
\end{figure}


{\em Proof.} We can rewrite ${\cal S}$ as
\begin{equation}
\begin{split}
\label{S}
\sum_{i \in V(G)} P(1|i) - |E(G)| + \sum_{(i,j) \in E(G)} \left[P(0,0|i,j) \right.\\
\left. +P(0,1|i,j)+P(1,0|i,j)\right] \equiv {\cal S'} - |E(G)|,
\end{split}
\end{equation}
where $|E(G)|$ is the cardinal of $E(G)$. Then,
\begin{equation}
{\cal S'} \stackrel{\mbox{\tiny{NCHVTs}}}{\leq} \alpha(G') \stackrel{\mbox{\tiny{QT}}}{\leq} \vartheta(G'),
\end{equation}
where $G'$ is the graph of the events in ${\cal S'}$. Therefore,
\begin{equation}
{\cal S} \stackrel{\mbox{\tiny{NCHVTs}}}{\leq} \alpha(G') - |E(G)| \stackrel{\mbox{\tiny{QT}}}{\leq} \vartheta(G') - |E(G)|.
\end{equation}
Now we have to prove that $\alpha(G')=\alpha(G)+|E(G)|$ and $\vartheta(G')=\vartheta(G)+|E(G)|$.

The maximum of ${\cal S'}$ for any NCHVT is always attained by assigning probability $1$ to some events and $0$ to the others. For every $(i,j) \in E(G)$, there are five vertices in $G'$: those corresponding to the probabilities of the events $1|i$, $1|j$, $0,0|i,j$, $0,1|i,j$, and $1,0|i,j$. See Fig.\ \ref{Fig2} (b). If $P(1|i)=1$ and $P(1|j)=1$, then $P(0,0|i,j)=0$, $P(0,1|i,j)=0$, and $P(1,0|i,j)=0$. Thus if we do this assignment for all edges in $G$ we obtain ${\cal S'}=|V(G)|$ (i.e., the number of vertices of $G$), which is less than or equal to $\alpha(G)+|E(G)|$. On the other hand, if $P(1|i)=0$ and $P(1|j)=0$, then $P(0,0|i,j)=1$, $P(0,1|i,j)=0$, and $P(1,0|i,j)=0$. Thus if we do this assignment for all edges in $G$, we obtain ${\cal S'}=|E(G)|$, which is strictly smaller than $\alpha(G)+|E(G)|$. If $P(1|i)=1$ and $P(1|j)=0$, then $P(0,0|i,j)=0$, $P(0,1|i,j)=0$, and $P(1,0|i,j)=1$. However, in general, we cannot do this assignment for all edges in $G$. An assignment maximizing ${\cal S'}$ for NCHVTs is one maximizing ${\cal S}$ (recall that ${\cal S'} = {\cal S} + |E(G)|$) and assigning probability $1$ to at most one of the vertices that are adjacent in $G$. This is exactly accomplished by any assignment leading to ${\cal S}=\alpha(G)$. Then, for every edge $(i,j)$ in $G$ with probabilities $1$ and $0$, the sum of the probabilities of the corresponding five-vertex subgraph of $G'$ is $2$, while for every $(i,j) \in E(G)$ with probabilities both $0$, the sum of the probabilities of the corresponding five-vertex subgraph of $G'$ is $1$. Therefore, the sum of the probabilities of the events in $G'$ is $\alpha(G)+|E(G)|$. This proves the upper bound of ${\cal S}$ for NCHVTs.

For the maximum of ${\cal S'}$ in QT consider the assignment of probabilities to the vertices of $G'$ corresponding, via Born's rule, to a Lov\'asz-optimum-orthogonal representation of $\overline{G}$. Since, for every edge $(i,j)$ of $G$, $P(1|i, 1|j)=0$, then $P(0,0|i,j)+P(0,1|i,j)+P(1,0|i,j)=1$. To prove that this is actually the quantum maximum notice that ${\cal S}$ (and ${\cal S'}$) contains only probabilities which are well defined within QT (irrespective of the order in which measurements are performed) only if all pairs $(i,j) \in E(G)$ correspond to compatible observables. Then, maximizing $\sum_{i\in V} |\langle i | \psi \rangle|^2-\sum_{(i,j) \in E(G)} | \langle j | i \rangle \langle i | \psi \rangle |^2$ is equivalent to maximizing $\sum_{i\in V(G)} |\langle i | \psi \rangle|^2$ under the restriction that $\langle j | i \rangle=0$ for $(i,j) \in E(G)$. Consequently, $\vartheta(G')=\vartheta(G)+|E(G)|$. \hfill \endproof

For clarity's sake we have not addressed the case of vertex-weighted graphs. However, the result presented in this section can be extended to vertex-weighted graphs with weights that are natural numbers (i.e., to the graphs corresponding to inequalities with rational coefficients) by noticing that any of these graphs can be converted into a standard graph by copying $n$ times each vertex with weight $n$.


\section{Discussion}


The observation that any NC inequality can be converted into a NC inequality of the form (\ref{main}) may change the way contextuality is tested in the laboratory. This is so because testing (\ref{main}) only requires measuring sequences of two observables, which allows us to simultaneously overcome many obstacles that, so far, had prevented experiments on most forms of contextuality. More specifically:

(i) On one hand, the fact that only two sequential measurements are required will help to improve the quality of the experimental results by reducing the imperfections and making easier to identify and correct them.

(ii) On the other hand, the fact that only two-point correlations are required allows us to apply existing tools for analyzing contextuality experiments with imperfections which are only defined in this case \cite{DKL15}.

(iii) Furthermore, it allows us to design tests free of problems (I) and (II) using systems in which sequential nondemolition measurements are not feasible. If only sequences of two measurements are needed, then the complexity of setup for testing each sequence is constant and thus problem (I) is solved. On the other hand, the first measurement can be simulated by a demolition measurement followed by a preparation that depends on the outcome. For example, the projective measurement of $i$ in Fig.\ \ref{Fig1} (b) can be simulated by a demolition measurement of $i$ followed by a preparation of a new system in state $|i\rangle$ if the outcome of $i$ is $1$, or in the state predicted by L\"uders's rule for the state $|\psi\rangle$ if the outcome of $i$ is $0$; see Fig.\ \ref{Fig1} (c). Actually, only the first part is needed for obtaining the probabilities in EQ.\ (\ref{main}), which can then be expressed as $P_{|\psi\rangle}(1,1|i,j)=P_{|\psi\rangle}(1|i) P_{|i\rangle}(1|j)$, where $P_{|\phi\rangle}(\ldots)$ denotes the probability $P_{|\phi\rangle}(\ldots)$ for the quantum state $|\phi\rangle$. Then, the choice of the second measurement is made after the result of the first measurement is indelibly recorded, thus solving problem (II).

Nevertheless, it is important to point out that the experiments using the scheme in Fig.\ \ref{Fig1} (c) present some conceptual disadvantages with respect to those using the schemes in Figs.\ \ref{Fig1} (a) and \ref{Fig1} (b). For example, the scheme in Fig.\ \ref{Fig1} (c) provides an operational justification for the assumption of outcome noncontextuality, but only in experiments with sequences of two measurements on $|\psi\rangle$. There, it is possible to check that, when $|\psi\rangle$ is prepared, $i$ does not affect the outcome probabilities for any $j$ that is measured afterwards. However, this is not guaranteed for other possible preparations because the way the measurement of $i$ in Fig.\ \ref{Fig1} (c) transforms the state depends on $|\psi\rangle$ (and, in particular, fails to imitate a projective measurement when the outcome is $i=0$). In contrast, experiments using the schemes in Figs.\ \ref{Fig1} (a) and \ref{Fig1} (b) provide a much more compelling operational justification of the assumption of outcome noncontextuality, since they allow us to check repeatability and no signaling for any preparation and any sequence of compatible measurements.

Notice also that the experiments in Figs.\ \ref{Fig1} (b) and \ref{Fig1} (c) are not thermodynamically equivalent. The measurement of $i$ in Fig.\ \ref{Fig1} (c), a demolition measurement followed by a preparation, dissipates more heat than the projective measurement of $i$ in Fig.\ \ref{Fig1} (b). Since classical simulations of quantum contextuality can be distinguished from truly quantum contextuality by the fact that the former produces more heat \cite{CGGLW15}, then to certify quantum contextuality it is more convenient to use tests like those in Fig.\ \ref{Fig1} (b) than those in Fig.\ \ref{Fig1} (c).

In addition to the advantages for designing better experiments, the fact that any form of contextuality can be revealed in experiments measuring only two compatible observables also has some conceptual advantages:

(iv) It allows for a novel black box approach to contextuality, extending the black box approach to nonlocality \cite{NGHA15}. The basic elements would be pairs of boxes with respective inputs $x,y \in V(G)$ such that $(x,y) \in E(G)$ and respective outputs $a,b\{0,1\}$. Unlike the case of nonlocality, here we are not limited by the condition that each box should be attributable to a party spacelike separated from the others. Still, we can use tools like postselection, composition, and wiring of boxes. This approach may help to understand why certain sets of nonlocal correlations \cite{NGHA15} are not realizable.

(v) It suggests a way for developing a resource theory of contextuality in analogy to the resource theory of nonlocality \cite{deVicente14}.

(vi) It permits translating any NC inequality into a communication complexity scenario involving only two-point communication. This may help in elucidating the connection between the principle that constraints quantum contextuality \cite{Cabello13} and information-theoretic principles that are essentially bipartite \cite{PPKSWZ09}.

(vii) It allows for building theory-independent sets of events with exclusivity relations leading to noncontextual and quantum bounds on demand. For any graph $G$, one can assign unit vectors to the vertices of $G$ such that adjacent vectors are orthogonal. This means that, according to QT, adjacent vertices correspond to alternative outcomes of a sharp measurement and, therefore, to exclusive events. However, these exclusivities cannot be justified in a theory-independent way unless we identify each of the vertices in $G$ with an operational procedure that explains why adjacent vertices correspond to exclusive events. The construction used in inequality (\ref{main}) avoids the problem of finding such procedures and allows us to construct theory-independent sets of events with noncontextual and quantum bounds equal to those of $G$. Notice that the only assumption in the transition from $G$ to $G'$ is that edges in $G$ correspond to jointly measurable observables.

(viii) The recent observation that contextuality is a necessary resource for fault-tolerant universal quantum computation via magic states \cite{HWVE14,DGBR15,RBDOB15} depends on a definition of contextuality based on the violation of inequalities involving probabilities of noncompatible observables (following Ref.\ \cite{CSW14}). The fact that any of these inequalities can be converted into a NC inequality of the type (\ref{main}) allows us to reformulate the contextuality-computation connection in terms of the traditional definition of contextuality as the violation of inequalities involving probabilities of compatible observables \cite{Peres02}.


\begin{acknowledgments}
We thank L.\ Aolita, M.\ Ara\'ujo, G.\ Ca\~{n}as, J.\ Chen, V.\ D'Ambrosio, E.\ S.\ G\'omez, J.-\AA.\ Larsson, G.\ Lima, A.\ J.\ L\'opez-Tarrida, S.\ P\'adua, J.\ R.\ Portillo, F.\ Sciarrino, S.\ van Dam, and A.\ Winter for useful conversations. This work was supported by the FQXi large grant project ``The Nature of Information in Sequential Quantum Measurements'' and Project No.\ FIS2014-60843-P, ``Advanced Quantum Information'' (MINECO, Spain), with FEDER funds.
\end{acknowledgments}



\end{document}